# Updating DMD Operators for Changes in Domain Properties

Dimitrios Voulanas[1,2*], Eduardo Gildin[1]

[1] Harold Vance Department of Petroleum Engineering, Texas A&M University, College Station, TX, USA

[2] Texas A&M Energy Institute, Texas A&M University, College Station, TX, USA

dvoulanas@tamu.edu, egildin@tamu.edu

*corresponding author

Abstract

Fast and reliable surrogate models are critical for optimization, control and uncertainty analysis in geological carbon-storage projects, yet high-fidelity multiphase simulators remain too expensive. Dynamic Mode Decomposition (DMD) offers an attractive data-driven reduction framework, but its operators are trained for a single set of reservoir properties. When permeability or well location changes, conventional practice is to regenerate snapshots and retrain the surrogate, erasing most of the speed advantage. This work presents a lightweight alternative that updates an existing DMD or DMD-with-control model without incorporating new simulation data or retraining. Two complementary update strategies are introduced. For cases where permeability changes uniformly across the domain, the proposed updates adjust the model's internal dynamics and control response to match the new flow timescale. When permeability varies in space, the approach modifies the spatial representation so that high-permeability zones are given greater influence on the model's reduced basis. Numerical experiments demonstrate that the proposed updates recover plume migration and pressure build-up within three percent of a freshly trained surrogate yet execute hundreds of times faster than full retraining. These methods therefore enable real-time optimization and rapid what-if studies while preserving the physical fidelity demanded by carbon-storage workflows.

## 1. Introduction

### 1.1 Overview

The management and optimization of subsurface resources, including geological carbon sequestration, enhanced oil recovery, and groundwater remediation, increasingly rely on sophisticated numerical models [1], [2]. These numerical simulations offer detailed predictions of fluid flow, geochemical interactions, and geomechanical responses, but their computational complexity poses a significant barrier for tasks such as real-time decision-making, uncertainty quantification, and optimization, which typically require numerous simulations [3], [4]. To address these computational challenges, data-driven surrogate and reduced-order models (ROMs) have become crucial. They approximate high-fidelity simulations with significantly lower computational demands, enabling extensive scenario analyses and optimization tasks [5].

ROM methods such as Proper Orthogonal Decomposition (POD), Trajectory Piecewise Linearization (TPWL), Dynamic Mode Decomposition (DMD), and other projection-based ROMs have been widely adopted in subsurface flow simulations. POD, for instance, identifies dominant spatial patterns or modes from high-resolution snapshots, significantly reducing the dimensionality of the problem [6], [5] (Bao et al., 2019; Jin, 2015). DMD, originally developed in fluid dynamics, has proven effective for subsurface flows, capturing coherent spatiotemporal modes from high-dimensional datasets [7]. Variants like DMD with control (DMDc) further enhance applicability by handling limited observations and external inputs, crucial for subsurface control problems [6], [8].

Concurrently, machine learning-based surrogates, particularly neural network models, Gaussian processes (GP), and hybrid physics-informed neural networks (PINNs), have increasingly been applied to subsurface simulation tasks [1], [4], [9]. Artificial Neural Networks (ANNs) and deep convolutional neural networks can efficiently approximate complex, nonlinear relationships between inputs (e.g., geological properties, injection rates) and outputs (e.g., pressures, saturation fields). These data-driven methods are particularly powerful due to their flexibility, although they require large training datasets [4]. Gaussian process emulators provide probabilistic predictions and uncertainty quantification, making them particularly suited for Bayesian optimization and sensitivity analyses in $CO_2$ sequestration.

The integration of these surrogate and reduced-order models into optimization workflows, such as design optimization, control, and history matching, has demonstrated considerable computational advantages. For example, the optimization of $CO_2$ injection well placement and rate schedules has significantly benefited from surrogate modeling, drastically reducing computational time while maintaining high accuracy in predictions [1]. Sun et al. (2021) demonstrated a workflow using a neural-network surrogate to optimize the location of $CO_2$ injection wells in a faulted saline reservoir. Their approach coupled a geomechanical reservoir simulator with an ANN that rapidly predicted outcomes (e.g. fault stability and storage volume) for different well locations [1]. Similarly, surrogate-assisted reinforcement learning approaches have been utilized for optimal pressure management during geological $CO_2$ sequestration, enabling real-time adaptive strategies that would otherwise be computationally infeasible. Chen et al. (2024) trained a deep surrogate using an encoder–transition–decoder network to learn the reservoir's pressure and plume dynamics in a latent reduced space. This made it feasible to train the reinforcement learning policy without thousands of expensive high-fidelity runs. The optimized strategy found by the agent significantly improved economic outcomes (e.g. net storage benefit) while keeping pressures below critical thresholds, outperforming baseline injection plans [10]. Srinivasan et al. (2021) presented a workflow for rapid forecasting and history matching in unconventional shale reservoirs using a combination of reduced-order decline-curve models and ANNs. In their approach, a fast but approximate physics model was used to generate abundant synthetic data, which a neural network learned from. That network was then fine-tuned with a small number of high-fidelity simulation runs to adapt it to the specific reservoir. The result was a physics-informed surrogate that could forecast production and also perform real-time history matching by updating the model as new data became available [9].

However, a significant challenge arises when the surrogate models face domain changes, such as variations in permeability distributions, operational parameters, or boundary conditions. Surrogates, especially those trained on limited or specific scenarios, may experience degraded predictive performance when conditions deviate significantly from their training datasets [6], [9]. Thus, maintaining or updating surrogate models to remain reliable under new conditions becomes crucial. Retraining surrogates entirely from scratch can negate the computational savings they originally provided. Therefore, methods that allow efficient updating or adapting surrogate models to new scenarios are critically needed.

Several strategies have been proposed to address surrogate updating, including incremental learning, transfer learning, and parametric ROMs. Incremental learning, for instance, enables surrogate models to incorporate new data sequentially, adapting to changing conditions without exhaustive retraining [9]. Transfer learning leverages previously trained surrogate models on similar scenarios, fine-tuning them with limited new data to achieve robust performance quickly [3]. Parametric ROMs explicitly incorporate varying parameters like permeability or porosity fields into the surrogate input, allowing them to generalize better to changes within these parameter spaces [6]. For example, one could include rock permeability fields or fluid property variations in the input vector to a surrogate model. Jin and

Durlofsky 2018 extended their POD-TPWL model to allow slight changes in the reservoir transmissibility by treating a scaling factor as a parameter. Their preliminary results showed the ROM could capture the general trends in well responses when all permeability values were scaled by a constant factor relative to the training case. However, the differences introduced were small; essentially the method worked for moderate perturbations but would likely need new training data for more drastic heterogeneity changes [11]. In general, covering a broad range of geologic variations (e.g. different facies distributions or fault configurations) requires either a very extensive training set (many simulations sampling those variations) or a means to update the model on the fly.

The successful updating of surrogate models relies significantly on the design of their training and adaptation strategies. For example, deep convolutional autoencoder models have been successfully combined with linear transition predictors in reduced latent spaces, providing efficient gradient computations for adjoint-based history matching in geological $CO_2$ sequestration scenarios [3]. Such hybrid models demonstrate robustness against domain changes, offering a computationally efficient means of surrogate updating and adaptability.

Despite these advancements, several practical and theoretical challenges remain. Key among them are data scarcity, computational cost of retraining, and the delicate balance between generalization and specialization. Data scarcity is particularly acute in subsurface modeling, where generating high-quality simulation data is resource-intensive and time-consuming [4]. Moreover, retraining or fine-tuning complex surrogate models, such as deep neural networks, demands significant computational resources, which may not be readily available in all operational settings [9]. Furthermore, ensuring that updated surrogates preserve physical consistency and accuracy, particularly when extrapolating beyond training scenarios/data, continues to be challenging [1], [6].

### 1.2 Aim of the Study

Given the outlined challenges and opportunities, the primary aim of this study is to develop and validate efficient methodologies for updating DMD-based surrogate models in subsurface porous media flow simulation, specifically within the context of $CO_2$ sequestration scenarios. The objectives of this study are:

1. Develop algebraic and numerical approaches to update existing DMD models efficiently without requiring extensive new simulations or retraining.

2. Quantify the accuracy, robustness, and computational efficiency of these updating methods through rigorous numerical experiments and comparison with traditional retraining approaches, where applicable.

3. Demonstrate the practical applicability of these updated surrogate models in realistic optimization and control problems encountered in $CO_2$ sequestration, including well placement optimization, injection scheduling, and pressure management.

By achieving these aims, the study intends to significantly enhance the utility and reliability of surrogate modeling techniques in subsurface engineering applications, providing practical solutions to the critical issue of surrogate model adaptability.

### 2. Materials and Methods

In this subsection we develop a coherent framework for uniform permeability scaling at both the continuum and reduced-order levels. Subsection 2.1.1 shows how a uniform change of permeability can be absorbed into a rescaling of time together with a consistent correction of the pressure amplitude, while leaving the saturation dynamics invariant. Subsection 2.1.2 analyses the induced transformation of the linearized generator and its discrete-time propagator, clarifying how eigenvalues and characteristic time scales are modified by the scaling. Subsection 2.1.3 addresses the practical issue of reconstructing surrogate snapshots at the original physical sampling times when the rescaled time step no longer coincides with the training output grid. Finally, Subsection 2.1.4 translates these results into algebraic update rules for the reduced-order operators associated with dynamic mode decomposition, so that permeability changes can be incorporated directly at the surrogate level without regenerating high-fidelity simulations or retraining the model.

## 2.1 Uniform Permeability Scaling

In multiphase reservoir simulation, the relationship between permeability and fluid transport speed is direct through Darcy's law. Consider the semi-discrete form of the pressure equation for slightly compressible, two-phase flow:

$$C\dot{p} - \nabla(K\Lambda(S)\nabla p) = q_p \tag{1}$$

where $C$ is accumulation coefficient (porosity × compressibility), $K$ is the absolute permeability, $\Lambda(S)$ is the total mobility as a function of saturation. The saturation transport equation is

$$\Phi\dot{S} + \nabla(f(S)\mathbf{u}) = q_s \tag{2}$$

with velocity given by Darcy's law

$$\mathbf{u} = -K\Lambda(S)\nabla p \tag{3}$$

Spatial discretization (e.g., finite volume or finite element) yields the matrix form

$$M_p \dot{p} + L(K,S)p = q_p \tag{4}$$

$$M_S \dot{S} + T(S,K)p = q_s \tag{5}$$

where $M_p$ and $M_S$ are mass/accumulation matrices, $L$ is the transmissibility matrix, and $T$ is the transport coupling term. For a uniform permeability change $K_{new} = \kappa K$, transmissibilities scale proportionally: $L_{new} = \kappa L$.

### 2.1.1 Time Scaling and Amplitude Correction

Introducing a stretched time variable $\tau = \kappa t$ and defining a scaled pressure $p_{new}(t) = \kappa^{-1} p_{old}(\tau)$ makes the accumulation and diffusion terms algebraically identical to those in the original system evaluated at $\tau$.

This shows that:

- The new pressure field evolves $\kappa$ times faster than in the reference case.
- A multiplicative factor $\kappa^{-1}$ on pressure restores physical balance for compressible fluids.
- The velocity field is unchanged apart from the time rescaling $\mathbf{u}_{new}(t) = -\kappa K \Lambda \nabla(\kappa^{-1} p_{old}(\kappa t)) = -K\Lambda \nabla p_{old}(\kappa t) = \mathbf{u}_{old}(\kappa t)$, so that the saturation equation sees exactly the same velocity field, but at the accelerated time $\tau$.

Thus,

$$S_{new}(t) = S_{old}(\kappa t) \tag{6}$$

$$p_{new}(t) = \kappa^{-1} p_{old}(\kappa t) \tag{7}$$

which provides the continuum-level justification for both the accelerated dynamics and the pressure amplitude correction.

### 2.1.2 Discrete-time Eigenvalue Modification

Linearizing about a state x gives $\dot{x} = L x + B u$ (or $\dot{p} = - M^{-1}T p + M^{-1}q$), where $L$ is the Jacobian or continuous-time generator. For a uniform scaling $K \to \kappa K$,

$$L_{\text{new}} = \kappa L \tag{8}$$

Continuous-time eigenvalues scale as: $\lambda_i^{(\text{new})} = \kappa \lambda_i^{(\text{old})}$, so time constants shrink by $1/\kappa$, and decay/oscillation rates scale by $\kappa$.

In discrete time over a step $\Delta t$, the state transition matrix is $A = e^{\Delta t L}$. Under the new permeability field,

$$A_{\text{new}} = e^{\Delta t \kappa L} = (e^{\Delta t L})^\kappa = A^\kappa \tag{9}$$

Thus, each discrete eigenvalue $\mu_i$ is mapped to $\mu_i^\kappa$. Stable modes ($|\mu_i| < 1$) decay more quickly for $\kappa > 1$, consistent with the faster physical process.

### 2.1.3 Reconstructing snapshots at original timestamps

If the new time step $\Delta t^\star = \Delta t / \kappa$ does not align with the original output times, new saturation fields can be recovered via interpolation from the original surrogate. For example, linear interpolation gives

$$S^{\text{new}}(\mathbf{x}, t_n) := \frac{t_{m+1} - \hat{t}_n}{\Delta t} S^{\text{old}}(\mathbf{x}, t_m) + \frac{\hat{t}_n - t_m}{\Delta t} S^{\text{old}}(\mathbf{x}, t_{m+1}) \tag{10}$$

where $\hat{t}_n = \kappa t_n$ and $t_m$ and $t_{m+1}$ bracket $\hat{t}_n$. Higher-order schemes (e.g., cubic Hermite) can improve temporal accuracy.

For pressure, post-multiplication by $\kappa^{-1}$ is applied after reconstruction to enforce the correct amplitude scaling.

### 2.1.4 Discrete Operators Update

When $k \in \mathbb{N}$, the updated DMDc operators are:

$$A_\kappa = A^\kappa \tag{11}$$

$$B_\kappa = = \sum_{j=0}^{\kappa-1} A^j B = (A^\kappa - I)(A - I)^{-1} B \tag{12}$$

When $k \in \mathbb{R}$, $A^\kappa$ is computed via a real Schur ($A = V \Lambda V^{-1}$), applying $\lambda_i^\kappa$ elementwise on the (block-)diagonal. The input matrix update uses the analytic continuation of the geometric sum:

$$\phi_\kappa(\lambda) = \begin{cases} \dfrac{\lambda^\kappa - 1}{\lambda - 1}, & |\lambda - 1| > \varepsilon \\ \kappa, & |\lambda - 1| \leq \varepsilon \end{cases} \tag{13}$$

with $B_\kappa = V \operatorname{diag}(\phi_\kappa(\lambda_i)) V^{-1} B$ for $A = V \operatorname{diag}(\lambda_i) V^{-1}$. Near-unit eigenvalues are handled by the $\kappa$-limit to avoid numerical instability. In practice, $(A - I)^{-1}$ should not be formed explicitly; the equation $(A - I)X = (A^\kappa - I)B$ should be solved with a linear solver.

This procedure yields updated surrogate operators that reproduce the dynamics of the modified permeability field at the original output timestamps, without regenerating training data or retraining the model.

## 2.2 Anisotropic Permeability Scaling

This sub-section extends the preceding analysis to anisotropic permeability fields by embedding the spatial variability of the rock fabric into a coordinate warp and the associated reduced operators. Subsection 2.2.1 introduces a permeability-conformal mapping that redistributes the snapshot degrees of freedom so that regions of large permeability occupy proportionally more volume in the warped grid while preserving appropriate energy norms. Subsection 2.2.2 refines this construction by defining an anisotropic warp and a local stretch metric that aligns the deformation with preferential-flow directions inferred from the permeability tensor, thereby distinguishing channel-aligned and cross-flow directions. Finally, Subsection 2.2.3 translates these geometric constructions into concrete update rules for the DMD operators, separating purely spatial rescaling from temporal rescaling and then combining them into a consistent framework for jointly modifying spatial bases and time propagators.

A snapshot matrix

$$X = [x_1, \ldots, x_{N_t}] \in \mathbb{R}^{n \times N_t}, \qquad n = n_x n_y n_z \tag{14}$$

assembled from fields $f(\mathbf{x}, t)$ on a uniform $(x, y, z)$-grid, is factorized by DMD through an unweighted Frobenius-norm minimization,

$$\min_{\Phi \in \mathbb{R}^{n \times r}, A_r, B_r} \sum_{x,y,z,t} |f(x,y,z,t) - f_{\text{DMD}}(x,y,z,t)|^2 \tag{15}$$

Because every cell contributes one summand, a permeable channel that covers a few cells may consume one mode, even though hydraulically it governs large-scale flow. Our goal here is to redistribute modal degrees-of-freedom (DoF) so that regions of large permeability K(x) are represented by proportionally more DoF.

### 2.2.1 Permeability-conformal coordinate warp

#### 2.2.1.1 Continuous formulation

Let $\Omega \subset \mathbb{R}^3$ denote the physical domain and a diffeomorphism $W: \Omega \to \tilde{\Omega}, (x, y, z) \mapsto (\xi, \eta, \theta)$ whose Jacobian determinant equals a power of permeability

$$|\det \nabla W| = K^\alpha(x, y, z), \quad \alpha \in \mathbb{R} \tag{16}$$

High-$K$ cells are therefore mapped to larger bricks in $(\xi, \eta, \theta)$-space, effectively multiplying their pixel count.

#### 2.2.1.2 Energy identity

For any scalar fields $f, g$ define $\tilde{f} = f \circ W^{-1}, \tilde{g} = g \circ W^{-1}$.

Change of variables yields

$$\iiint_{\tilde{\Omega}} \tilde{f}\, \tilde{g}\, d\xi\, d\eta\, d\theta = \iiint_{\Omega} f\, g\, K^\alpha dx\, dy\, dz \tag{17}$$

Performing SVD or DMD on $\tilde{f}$ is equivalent to performing a $K^\alpha$-weighted SVD on the original grid plus a geometric stretch that multiplies the local DoF.

$$\min_{\Phi\in\mathbb{R}^{n\times r},A_r,B_r} \sum_{x,y,z,t} K^\alpha(x,y,z)\,|\,f(x,y,z,t) - f_{DMD}(x,y,z,t)\,|^2 \tag{18}$$

Because $K(x,y,z)$ is sampled on a tensor grid with centers $(x_i, y_j, z_k)$ we can build a separable map via three one-dimensional cumulative integrals

$$\xi(x) = C_x \int_{x_0}^{x} \bar{K}^\alpha_x(s)\,ds \tag{19}$$

$$\eta(y) = C_y \int_{y_0}^{y} \bar{K}^\alpha_y(s)\,ds \tag{20}$$

$$\theta(z) = C_z \int_{z_0}^{z} \bar{K}^\alpha_z(s)\,ds \tag{21}$$

where $\bar{K}^\alpha_x(s)$ is the cross-sectional mean of $K^\alpha$ over $y, z$ at abscissa $s$ (and analogously for $y$ and $z$). The constants $C_{x,y,z}$ are chosen so that $(\xi, \eta, \theta)$ range over the grid indices $[0, n_x - 1]$, $[0, n_y - 1]$, $[0, n_z - 1]$, thereby preserving array shape.

The discrete Jacobian per cell, with incremental stretches, is:

$$J_{ijk} = \frac{\Delta\xi_j \Delta\eta_i \Delta\theta_k}{\Delta x\,\Delta y\,\Delta z} \approx K^\alpha_{ijk} \tag{22}$$

where $\Delta\xi_j = \xi_{j+1} - \xi_j$, $\Delta\eta_i = \eta_{i+1} - \eta_i$, $\Delta\theta_k = \theta_{k+1} - \theta_k$ are the finite differences

$$\sum_{ijk} (S^{-1}f)_{ijk}\,(S^{-1}g)_{ijk} = \sum_{ijk} f_{ijk} g_{ijk} J_{ijk} \tag{23}$$

hence $S^{-1}$ applied to snapshots mimics the warp's $K^\alpha$ weight. Also, $S$ is the diagonal matrix with entries $S_{ijk} = J_{ijk}^{-1/2}$, so $S^{-1}$ scales each snapshot component by $J_{ijk}^{1/2}$.

### 2.2.1.3 Snapshot resampling on the stretched grid

Each original snapshot $f(\cdot, t)$, typically defined on a Cartesian tensor product grid in $(x, y, z)$, is interpolated onto the new regular stretched grid. This interpolation is carried out by constructing a trilinear or tricubic interpolant $I_t(x, y, z)$ from the values of the snapshot at grid points. For every stretched-grid cells located at coordinates $(\xi_i, \eta_j, \theta_k)$, its value in the warped snapshot is given by:

$$\tilde{f}_{ijk}(t) = I_t(\xi_i, \eta_j, \theta_k) \tag{24}$$

yielding the warped snapshot $\tilde{x}_t \in \mathbb{R}^n$, where $n = n_x n_y n_z$, and a full snapshot matrix $\tilde{X} = [\tilde{x}_1, \ldots, \tilde{x}_{N_t}]$. Since the warped grid is stretched in regions of high permeability, such regions contribute more entries to the snapshot matrix and hence receive more emphasis during subsequent DMD processing. However, this interpolation process preserves only the local function shape - not integral quantities like total $K^\alpha$-weighted saturation.

If strict conservation of total $K^\alpha$-weighted saturation is needed, the interpolated field must be replaced by a conservative resampling that guarantees integral preservation of the transported scalar (e.g., saturation, water volume).

Let $S(x, y, z, t)$ be the water saturation at time $t$ defined on the original grid. Suppose that for a given cells in the stretched grid, the warped coordinates $(\xi_i, \eta_j, \theta_k)$ define the corner of a deformed brick in physical space that corresponds approximately to some nonuniform "old" cell in the original $(x, y, z)$-grid. The goal is to define a resampled value $\tilde{S}_{ijk}(t)$ that, when multiplied by the volume of the new cell in stretched coordinates (given by the Jacobian determinant), preserves the total weighted saturation in the original cell.

To this end, we define:

$$\tilde{S}_{ijk}(t) = \frac{1}{J_{ijk}} \int_{\text{cell}_{ijk}} S(x, y, z, t) \cdot K^\alpha(x, y, z) \, dx \, dy \, dz \tag{25}$$

where $\text{cell}_{ijk}$ denotes the region in the original domain that maps to the cell centered at $(\xi_i, \eta_j, \theta_k)$ and is defined by $W^{-1}(\tilde{C}_{ijk})$ with $\tilde{C}_{ijk} = [\xi_j, \xi_{j+1}] \times [\eta_i, \eta_{i+1}] \times [\theta_k, \theta_{k+1}]$, $J_{ijk}$ is the discrete approximation of $|\det \nabla W| \approx K^\alpha_{ijk}$ for that cell, and the division by $J_{ijk}$ ensures the result is a local cell-averaged saturation in the stretched space.

This ensures that:

$$\sum_{i,j,k} J_{ijk} \cdot \tilde{S}_{ijk}(t) = \sum_{i,j,k} \int_{\text{cell}_{ijk}} S(x, y, z, t) \cdot K^\alpha(x, y, z) \, dx \, dy \, dz$$

$$= \int_\Omega S(x, y, z, t) \cdot K^\alpha(x, y, z) \, dV \tag{26}$$

which is the total $K^\alpha$-weighted saturation in the physical domain.

This conservative interpolation modifies only the resampling step. Once the warp-space snapshots $\tilde{x}_t$ are constructed using these conservative $\tilde{S}_{ijk}(t)$, the DMD algorithm proceeds identically: computing a low-rank factorization $\tilde{X} \approx \tilde{\Phi} \tilde{S} \tilde{V}^\top$, forming the reduced dynamics matrix $\tilde{A} = \tilde{\Phi}^\dagger \tilde{X}' \tilde{V} \tilde{S}^{-1}$, and mapping the modes $\tilde{\phi}^{(\ell)}$ back to physical coordinates via the inverse warp as shown below. Because all subsequent operations are performed in warp space, enforcing Eq. (26) once per snapshot guarantees that every low-rank reconstruction obtained from $\tilde{\Phi}, \tilde{A}, \tilde{B}$ preserves the same weighted integral when pulled back to physical coordinates.

Because the Frobenius norm is unweighted in $(\xi, \eta, \theta)$, Eq. (13) guarantees this is a $K^\alpha$-weighted fit in physical space. High-$K$ regions, now occupying more grid nodes, force the low-rank basis to assign them additional modes.

### 2.2.1.4 Mapping modes back to the physical grid

Each warp-space mode $\tilde{\phi}^{(\ell)}_{ijk}$ corresponds to a continuous field $\tilde{\phi}^{(\ell)}(\xi, \eta, \theta)$ (piecewise trilinear). Define the physical mode via pull-back:

$$\phi^{(\ell)}(x, y, z) = \tilde{\phi}^{(\ell)}\left(\xi(x), \eta(y), \theta(z)\right) \tag{27}$$

Stacking $\phi^{(\ell)}$ as columns gives $\Phi \in \mathbb{R}^{n \times r}$.

### 2.2.2 Definition and local stretch metric

#### 2.2.2.1 Anisotropic warp

The isotropic stretch of redistributes degrees-of-freedom proportionally to the magnitude of a scalar permeability field. When the rock fabric is direction-dependent, or when principal flow paths are oblique to the Cartesian axes, the modal budget must also honor orientation. We therefore generalize the warp to a full $3 \times 3$ Jacobian field.

Let the permeability be a symmetric positive-definite tensor

$$\mathbf{K}(x,y,z) \in \mathbb{R}^{3\times 3}, \qquad 0 < k_{\min} \leq \lambda_{\min}(\mathbf{K}) \leq \lambda_{\max}(\mathbf{K}) \leq k_{\max} < \infty \tag{28}$$

An anisotropic permeability-conformal warp is any $C^1$ diffeomorphism

$$W: \Omega \longrightarrow \tilde{\Omega}, (x,y,z) \longmapsto (\xi, \eta, \theta) \tag{29}$$

whose Jacobian matrix is $M(x,y,z) = \nabla W = \begin{bmatrix} \partial\xi/\partial x & \partial\xi/\partial y & \partial\xi/\partial z \\ \partial\eta/\partial x & \partial\eta/\partial y & \partial\eta/\partial z \\ \partial\theta/\partial x & \partial\theta/\partial y & \partial\theta/\partial z \end{bmatrix}$ as such to satisfy:

$$\det M(x,y,z) = (\det \mathbf{K}(x,y,z))^{\alpha/3}, \alpha \in \mathbb{R} \tag{30}$$

The local stretch anisotropy is

$$A(x,y,z) = \frac{\sigma_1}{\sigma_3} \geq 1 \tag{31}$$

equal to one in the isotropic limit. Volume inflation scales with $(\det \mathbf{K})^{\alpha/3}$, but the shape of each brick is now governed by the full tensor.

The energy identity here extends to:

$$\int_{\tilde{\Omega}} \tilde{f}\, \tilde{g}\, d\xi\, d\eta\, d\theta = \int_{\Omega} f\, g\, (\det \mathbf{K})^{\alpha/3} dV \tag{32}$$

so performing DMD on the warped snapshots minimises a $(\det K)^{\alpha/3}$-weighted error in physical space.

#### 2.2.2.2 Directional (streamline-aligned) integral warp

Let $d(x,y,z)$ be a unit vector field that tracks the principal flow direction (e.g. the dominant eigenvector of K). Introduce a curvilinear coordinate triple $(s, n_1, n_2)$ where $s$ is arclength along integral curves of $d$ and $n_1, n_2$ are orthogonal distances in the normal plane spanned by two-unit vectors $e_1, e_2 \perp d$.

Define

$$\xi(s) = C_\| \int_{s_0}^{s} (\det \mathbf{K})^{\alpha_\|/3}(s')\, ds' \tag{33}$$

$$\eta(n_1) = C_{\perp 1} \int_{0}^{n_1} (\det \mathbf{K})^{\alpha_\perp/3}(n_1')\, dn_1' \tag{34}$$

$$\theta(n_2) = C_{\perp 2} \int_0^{n_2} (\det \mathbf{K})^{\alpha_\perp/3}(n_2') \, dn_2' \qquad (35)$$

The constants are fixed by the range conditions $\xi \in [0, n_\xi - 1]$, $\eta \in [0, n_\eta - 1]$, $\theta \in [0, n_\theta - 1]$.

The Jacobian determinant is

$$\det M = C_\| C_{\perp 1} C_{\perp 2} \, (\det \mathbf{K})^{(\alpha_\| + 2\alpha_\perp)/3} \qquad (36)$$

Choosing $C_\| C_{\perp 1} C_{\perp 2} = 1$ and $\alpha_\| + 2\alpha_\perp = \alpha$ enforces (23). The aspect ratio of the stretched brick obeys $A \sim \exp\left(\frac{1}{3}(\alpha_\| - \alpha_\perp)\ln \kappa\right)$ with $\kappa = \lambda_{\max}(\mathbf{K})/\lambda_{\min}(\mathbf{K})$, so setting $\alpha_\| > \alpha_\perp$ lengthens bricks along the channel while leaving cross-flow dimensions nearly unchanged.

### 2.2.2.3 Preferential-Flow Direction Map in 3-D

A preferential-flow direction map is the mathematical bridge between the raw permeability data held on a three-dimensional reservoir grid and the anisotropic coordinate warp that will redistribute dynamic-mode-decomposition (DMD) degrees-of-freedom. The construction has three logical layers—first, a direction field $d(x)$; second a stretch tensor $M(x)$ that inflates space preferentially along $d$; and finally a weighted Laplace system whose solutions $(\xi, \eta, \theta)$ form the warp.

Let $\Omega \subset \mathbb{R}^3$ with coordinates $x = (x, y, z)$. At every point attach the symmetric positive-definite permeability tensor

$$\mathbf{K}(x) = \begin{bmatrix} K_{xx} & K_{xy} & K_{xz} \\ K_{yx} & K_{yy} & K_{yz} \\ K_{zx} & K_{zy} & K_{zz} \end{bmatrix} \qquad (37)$$

let $k_1 \geq k_2 \geq k_3 > 0$ be its eigenvalues and $e_1, e_2, e_3$ the corresponding orthonormal eigenvectors, so that

$$\mathbf{K} = e_1 k_1 e_1^\top + e_2 k_2 e_2^\top + e_3 k_3 e_3^\top \qquad (38)$$

The tensor itself encodes preferential flow orientation because a Darcy flux vector $q$ of fixed magnitude produces volumetric flux $q^\top \mathbf{K} q$. This quadratic form is maximised by choosing $q$ parallel to the largest eigenvector, so define

$$d(x) = \frac{e_{\max}(\mathbf{K}(x))}{\|e_{\max}\|} \qquad (39)$$

the unit eigenvector $e_{\max} = e_1$ belonging to $k_1$. It maximizes volumetric flux over unit vectors $q$ and is unique unless $k_1 = k_2$. When $k_1 = k_2$ the tensor is locally isotropic in the plane spanned by $e_1$ and $e_2$.

### 2.2.2.4 Anisotropic stretch tensor $M(x)$

A coordinate warp must satisfy two simultaneous goals. First, its Jacobian determinant must be a prescribed power of permeability so that the volume element in warp space captures regions of high transmissibility. Second, its shape must elongate blocks along the flow direction with an anisotropy ratio that can be tuned independently of the volumetric weighting.

Let a permeability exponent $\alpha \in \mathbb{R}$ and a local anisotropy ratio $A(x) \geq 1$ as such $A(x) = (k_1/k_3)^\beta$.

Let $\Delta(x) = \det \mathbf{K}(x) = k_1 k_2 k_3$

and

$$\sigma_\perp(x) = \Delta(x)^{\alpha/3} A(x)^{-1/3} \tag{40}$$

$$\sigma_\parallel(x) = \Delta(x)^{\alpha/3} A(x)^{2/3} \tag{41}$$

These choices satisfy $\sigma_\parallel/\sigma_\perp = A$ and $\sigma_\parallel \sigma_\perp^2 = \Delta^\alpha$. The stretch tensor

$$M(x) = \sigma_\parallel(x)\, d(x)d(x)^\top + \sigma_\perp(x)[I - d(x)d(x)^\top] \tag{42}$$

is symmetric positive-definite and obeys

$$\det M(x) = (\det \mathbf{K}(x))^\alpha \tag{43}$$

where $d$ is whichever direction field has been adopted. By construction $M > 0$ and $\det M = \Delta^\alpha$. The principal stretches are therefore $A$-times larger along $d$ than across it, yet the cell volume in warp space incorporates the desired permeability power. When $A = 1$ the warp reduces smoothly to the earlier isotropic case.

### 2.2.3 Discrete Operators Update

The inverse matrices are placed on the left because they act on the "row side". The direct matrices are places on the right because they act on the "column side". This makes the procedures below similarity transformations, so to keep the physics intact but change their response subject to different domain properties like permeability.

#### 2.2.3.1 Spatial Scaling

The spatial scaling factor $(SF = (sf_1, \ldots, sf_n) > 0)$ $(n \times n)$ is applied row-wise to the DMD components after being unprojected using $\Phi$.

$$A = \Phi \tilde{A} \Phi^\dagger \tag{44}$$

$$B = \Phi \tilde{B} \tag{45}$$

The spatial scaling factor is applied to the dense $A$ matrix and then project back to reduced form below:

$$A^{(sp)} = SF^{-1} A\, SF \tag{46}$$

$$\tilde{A}^{(sp)} = (\Phi^{(sp)})^\dagger A^{(sp)} \Phi^{(sp)} \tag{47}$$

where $\Phi^{(sp)} = SF\Phi$ and $(\Phi^{(sp)})^\dagger = \Phi^\dagger SF^{-1}$.

Regarding $B$ matrix, it is first unprojected using Eq. (45) then project back with $(\Phi^{(sp)})^\dagger$:

$$B^{(sp)} = (\Phi^\dagger SF^{-1})(\Phi B) \tag{48}$$

Forming $A$ explicitly, as shown above, requires high memory usage. Using Eqs. (49) and (50) modification of spatial basis $\Phi$ can be done without fully unprojecting the matrix to full state space.

$$\Phi^{(sp)*} = SF\, \Phi = sf \cdot \Phi \tag{49}$$

$$\Phi^{(sp)\dagger} = (SF\Phi)^\dagger = (sf \cdot \Phi)^\dagger \tag{50}$$

since

$$\Phi^{(sp)\dagger} = \Phi^{(sp)\dagger} \Phi^{(sp)} \Phi^{(sp)\dagger} = \left(\Phi^\dagger SF^{-1}\right)(SF\Phi)\left(\Phi^\dagger SF^{-1}\right) = \tag{51}$$

$$=\Phi^\dagger \Phi \Phi^\dagger SF^{-1} = \Phi^\dagger SF^{-1} = (SF\Phi)^\dagger$$

Finally, a more efficient way is to project the scaling factor to projected space using the original modes, $\Phi$, and their pseudo-inverse, $\Phi^\dagger$, as follows using Gram matrices. This approach keeps $\Phi$ and $\Phi^\dagger$ untouched so the scaling won't be counted twice. Also, these matrices are well-conditioned, semi-positive definite by construction, and compensate for loss of Euclidean orthogonality for a basis of full column rank.

$$G_L = \Phi^\dagger SF^{-1}\Phi \tag{52}$$

$$G_R = \Phi^\dagger SF\Phi \tag{53}$$

therefore

$$\tilde{A}^{(sp)} = \left(\Phi^\dagger SF^{-1}\Phi\right)\tilde{A}\left(\Phi^\dagger SF\Phi\right) = \Phi^\dagger SF^{-1}\left(\Phi\tilde{A}\Phi^\dagger\right)SF\Phi = \Phi^\dagger SF^{-1}ASF\Phi = \Phi^\dagger A^{(sp)}\Phi \tag{54}$$

$$\tilde{A}^{(sp)} = G_L \tilde{A} G_R \tag{55}$$

Regarding the $B$ matrix, it must be the same before and after the spatial scaling modification:

$$\Phi\tilde{B}^{(sp)} = \Phi G_L\tilde{B} \tag{56}$$

### 2.2.3.2 Temporal Scaling

The temporal scaling factor ($\Gamma = \text{diag}(\gamma_1, \ldots, \gamma_r) > 0$) ($r \times r$) is applied column-wise to the full state DMD components as shown below.

$$A^{(t)} = T^{-1}(\Phi\tilde{A}\Phi^\dagger)T = (\Phi\Gamma^{-1}\Phi^\dagger)(\Phi\tilde{A}\Phi^\dagger)(\Phi\Gamma\Phi^\dagger) = \Phi\left(\Gamma^{-1}I_r\,\tilde{A}I_r\Gamma\right)\Phi^\dagger = \tag{57}$$
$$= \Phi\,\tilde{A}^{(t)}\,\Phi^\dagger$$

Eq. (57) puts the temporal scaling factors entirely inside the reduced $\tilde{A}$, which should always be valid. Also, scaling both the $\Phi$ and the $\tilde{A}$ leads to the temporal scaling factors being canceled out: $\Phi\Gamma(\Gamma^{-1}\tilde{A}\Gamma)\Gamma^{-1}\Phi^\dagger = \Phi\tilde{A}\Phi^\dagger$.

$$B^{(t)} = \Phi\tilde{B} = T^{-1}B = \Gamma^{-1}\tilde{B} \tag{58}$$

where $T = \Phi\Gamma\Phi^\dagger$, $T^{-1} = \Phi\Gamma^{-1}\Phi^\dagger$, $T^{-1}\Phi = \Phi\Gamma^{-1}$ and $\Phi^\dagger T = \Gamma\Phi^\dagger$.

Notice, that if $\Phi$ is modified, then $\tilde{A}, \tilde{B}$ must remain unchanged. If $\tilde{A}, \tilde{B}$ are chosen to be modified, then $\Phi$ must remain unchanged same as in the spatial modification.

Directly in reduced space, for keeping $\Phi$ unchanged and modifying the $\tilde{A}, \tilde{B}$ matrices:

$$\tilde{A}^{(t)} = \Gamma^{-1}\tilde{A}\Gamma \tag{59}$$

Similarly, for the B matrix:

$$\tilde{B}^{(t)} = \Gamma^{-1}\tilde{B} \tag{60}$$

### 2.2.3.3 Spatial and Temporal Scaling

Applying both spatial and temporal scaling to the DMD components is denoted by the following formulas in full space state. Since there is no evidence of which transformation to apply first both must be explored. First, case A, apply the spatial scaling factor and then the temporal scaling factor:

$$\begin{aligned}A^{(sp,t)} &= T^{-1}SF^{-1}\left(\Phi\tilde{A}\Phi^\dagger\right)SFT = T^{-1}SF^{-1}A\,SFT \\ &= \Phi\,\Gamma^{-1}\,\Phi^\dagger\,SF^{-1}(\Phi\,\tilde{A}\Phi^\dagger)SF\,\Phi\,\Gamma\,\Phi^\dagger = \Phi\Gamma^{-1}G_L\,\tilde{A}\,G_R\Gamma\Phi^\dagger \\ &= \Phi\,\Gamma^{-1}\,\Phi^\dagger\,SF^{-1}ASF\,\Phi\,\Gamma\,\Phi^\dagger \end{aligned} \tag{61}$$

$$B^{(sp,t)} = T^{-1}\,SF^{-1}\Phi\,\tilde{B} = \left(\Phi\Gamma^{-1}\Phi^\dagger\right)SF^{-1}B = \Phi\,\Gamma^{-1}(\Phi^\dagger\,SF^{-1}\Phi\,)\tilde{B} = \Phi\,\Gamma^{-1}\,G_L\,\tilde{B} \tag{62}$$

Case B is to apply the temporal scaling factor and then the spatial scaling factor:

$$\begin{aligned}A^{(sp,t)} &= SF^{-1}T^{-1}(\Phi\,\tilde{A}\Phi^\dagger)T\,SF = SF^{-1}\Phi\,\Gamma^{-1}(\Phi^\dagger\Phi)\,\tilde{A}(\Phi^\dagger\Phi)\,\Gamma\,\Phi^\dagger SF \\ &= SF^{-1}\Phi\Gamma^{-1}I_r\,\tilde{A}\,I_r\Gamma\Phi^\dagger SF = SF^{-1}\Phi\Gamma^{-1}\,\tilde{A}\,\Gamma\Phi^\dagger SF = SF^{-1}\Phi\tilde{A}^{(t)}\Phi^\dagger SF \end{aligned} \tag{63}$$

$$\begin{aligned}B^{(sp,t)} &= SF^{-1}\,T^{-1}\,\Phi\,\tilde{B} = SF^{-1}\left(\Phi\Gamma^{-1}\Phi^\dagger\right)B = SF^{-1}\left(\Phi\Gamma^{-1}\Phi^\dagger\right)\Phi\tilde{B} = SF^{-1}\Phi\Gamma^{-1}I_r\tilde{B} \\ &= SF^{-1}\Phi\Gamma^{-1}\tilde{B} = SF^{-1}\Phi\tilde{B}^{(t)} = SF^{-1}B^{(t)}\end{aligned} \tag{64}$$

where $T = \Phi\,\Gamma\,\Phi^\dagger$, $T^{-1} = \Phi\,\Gamma^{-1}\,\Phi^\dagger$, $T^{-1}\Phi = \Phi\Gamma^{-1}$ and $\Phi^\dagger T = \Gamma\,\Phi^\dagger$.

However, this requires forming a large dense $A$ matrix. This can be avoided in two ways. The first way involves applying the scaling factors to operators $\tilde{A}, \tilde{B}$ while leaving $\Phi$ untouched. The second way leaves the $\tilde{A}$ operator untouched and scales the spatial basis, $\Phi$, and modifies the $\tilde{B}$ so it can stay the same as the dense B.

The first case applies the scaling factors to the operators.

Regarding operator A in case A approach:

$$A^{(sp,t)} = \Phi\tilde{A}^{(sp,t)}\Phi^\dagger = \Phi\Gamma^{-1}G_L\,\tilde{A}\,G_R\Gamma\Phi^\dagger = \Phi\Gamma^{-1}(\Phi^\dagger SF^{-1}\Phi)\,\tilde{A}\,(\Phi^\dagger SF\Phi)\Gamma\Phi^\dagger \Leftrightarrow \tag{65}$$

$$A^{(sp,t)} = \Phi\Gamma^{-1}\Phi^\dagger SF^{-1}\left(\Phi\tilde{A}\Phi^\dagger\right)SF\Phi\Gamma\Phi^\dagger = \Phi\Gamma^{-1}\Phi^\dagger SF^{-1}ASF\Phi\Gamma\Phi^\dagger = T^{-1}SF^{-1}ASFT \tag{66}$$

Case B approach:

$$\begin{aligned}A^{(sp,t)} &= \Phi\tilde{A}^{(sp,t)}\Phi^\dagger = \Phi G_L\,\Gamma^{-1}\tilde{A}\,\Gamma G_R\Phi^\dagger = \Phi(\Phi^\dagger SF^{-1}\Phi)\tilde{A}^{(t)}(\Phi^\dagger SF\Phi)\Phi^\dagger \\ &= \Phi\Phi^\dagger SF^{-1}A^{(t)}\,SF\Phi\Phi^\dagger \Leftrightarrow \end{aligned} \tag{67}$$

$$A^{(sp,t)} = I_n SF^{-1}A^{(t)}SFI_n = SF^{-1}A^{(t)}SF \tag{68}$$

Regarding operator B in case A approach:

$$B^{(sp,t)} = \Phi \tilde{B}^{(sp,t)} = \Phi \Gamma^{-1} G_L \tilde{B} = \Phi \tilde{B}^{(sp,t)} \Leftrightarrow \tag{69}$$

$$B^{(sp,t)} = (\Phi \Gamma^{-1} \Phi^\dagger SF^{-1} \Phi) \tilde{B} \Leftrightarrow \tag{70}$$

$$B^{(sp,t)} = (T^{-1} SF^{-1}) \Phi \tilde{B} \Leftrightarrow \tag{71}$$

$$B^{(sp,t)} = (T^{-1} SF^{-1}) B \tag{72}$$

Case B approach:

$$B^{(sp,t)} = \Phi \tilde{B}^{(sp,t)} = \Phi G_L \Gamma^{-1} \tilde{B} = \Phi \tilde{B}^{(sp,t)} \Leftrightarrow \tag{73}$$

$$B^{(sp,t)} = (SF^{-1} \Phi) \tilde{B}^{(t)} \Leftrightarrow \tag{74}$$

$$B^{(sp,t)} = (I_n SF^{-1}) \Phi \tilde{B}^{(t)} \Leftrightarrow \tag{75}$$

$$B^{(sp,t)} = SF^{-1} B^{(t)} \tag{76}$$

Case A approach is recommended because it does not lead the formation of $I_n$ matrix. This formation is only valid if $\Phi$ is square and of full rank. If not, then $\Phi \Phi^\dagger$ cannot be replaced by $I_n$. In this case, $P = \Phi \Phi^\dagger$ being a

The second way, while $\tilde{A}, \tilde{B}$ remain untouched, the scaled $\Phi$ and $\Phi^\dagger$ is given by:

$$\Phi^{(sp,t)} = SF\Phi\Gamma \tag{77}$$

$$\Phi^{(sp,t)\dagger} = (SF\Phi\Gamma)^\dagger = \Gamma^{-1} \Phi^\dagger SF^{-1} \tag{78}$$

Regarding the A operator, in this case:

$$\Phi^{(sp,t)} \tilde{A} \, a_k = SF \, \Phi \, \Gamma \, \tilde{A} \left( \Gamma^{-1} \Phi^\dagger SF^{-1} x_k \right) = SF \, \Phi \Gamma \tilde{A} \Gamma^{-1} \Phi^\dagger \, SF^{-1} x_k = \\ = SF \, \Phi \, \Gamma^{-1} \tilde{A} \Gamma \Phi^\dagger SF^{-1} x_k \tag{79}$$

Same as before the equality $\Gamma \tilde{A} \Gamma^{-1} = \Gamma^{-1} \tilde{A} \Gamma$ holds only if $\Gamma$ commutes with $\tilde{A}$ ($\tilde{A}\Gamma = \Gamma \tilde{A}$).

Since all scaling is kept in the basis, the following B operator equality must hold:

$$\Phi^{(sp,t)} \tilde{B}^{(sp,t)} = \Phi \tilde{B} \Leftrightarrow \tag{80}$$

$$SF\Phi\Gamma \tilde{B}^{(sp,t)} = \Phi \tilde{B} \Leftrightarrow \tag{81}$$

$$\Phi^\dagger SF\Phi\Gamma \tilde{B}^{(sp,t)} = \Phi^\dagger \Phi \tilde{B} = I_r \tilde{B} \Leftrightarrow \tag{82}$$

$$G_R \Gamma \tilde{B}^{(sp,t)} = \tilde{B} \Leftrightarrow \tag{83}$$

$$\tilde{B}^{(sp,t)} = \Gamma^{-1} G_R^{-1} \tilde{B} \tag{84}$$

Also,

$$B^{(sp,t)} = \Phi^{(sp,t)} \tilde{B}^{(sp,t)} = S_F \Phi \Gamma (\Gamma^{-1} G_R^{-1} \tilde{B}) = S_F \Phi G_R^{-1} \tilde{B} \tag{85}$$

because

$$S_F \Phi G_R^{-1} = S_F \Phi (\Phi^\dagger S_F \Phi)^{-1} = \Phi (\Phi^\dagger S_F \Phi)(\Phi^\dagger S_F \Phi)^{-1} = \Phi \tag{86}$$

Therefore, $B^{(sp,t)} = \Phi \tilde{B} = B$.

Putting altogether, in terms of the original spatial basis and reduced operators:

Regarding the dense matrix case,

$$x_{k+1} = SF^{-1} \Phi \Gamma^{-1} \tilde{A} \Gamma \Phi^\dagger SF x_k + SF^{-1} \Gamma^{-1} \Phi \tilde{B} u_k \tag{87}$$

For the spatial basis case,

$$x_{k+1} = SF \Phi \Gamma [\tilde{A}(\Gamma^{-1} \Phi^\dagger SF^{-1} x_k) + \Gamma^{-1} G_R^{-1} \tilde{B} u_k] \tag{88}$$

Regarding the Gram matrix approach,

$$x_{k+1} = \Phi [(\Gamma^{-1} G_L \tilde{A} G_R \Gamma)(\Phi^\dagger x_k) + (\Gamma^{-1} G_L \tilde{B}) u_k] \tag{89}$$

where $G_L = \Phi^\dagger SF^{-1} \Phi$ and $G_R = \Phi^\dagger SF \Phi$.

## 3. Results and Discussion

This section evaluates the proposed operator updates on a representative multiphase flow configuration, using permeability perturbations that range from modest to very pronounced. For each scenario, a DMD or DMD-with-control surrogate trained at the reference permeability is compared against two surrogates: one obtained by directly updating the reduced operators according to the scaling rules developed in Section 2, and one left unchanged and thus "non scaled." Errors are reported in terms of mean absolute error (MAE) and a complementary PCE metric, both computed over the full domain and over regions near and far from the injection well, so as to separate large-scale plume behaviour from the more nonlinear near-wellbore dynamics. Tables 1–4 summarize these quantitative results for uniform and non-uniform permeability changes.

### 3.1 Uniform Permeability Change

For uniform permeability multipliers, Tables 1 and 2 show a systematic advantage of the updated operators over the non-scaled baseline across all tested values. At the global level, the scaled surrogate reduces the pressure MAE by roughly two orders of magnitude when the permeability is reduced to forty percent of its reference value, and by a similar factor when it is increased slightly to ninety or one hundred and ten percent. The far-wellbore region exhibits an even stronger effect: errors there drop by three orders of magnitude when scaling is applied, indicating that the time-rescaling and amplitude correction correctly capture the accelerated or decelerated propagation of pressure signals through the bulk of the domain.

The near-wellbore region is more challenging, as expected, because of strong nonlinearities and steep gradients. Nevertheless, the scaled surrogate still lowers the MAE by roughly one order of magnitude for all uniform multipliers, including the extreme four-hundred-percent case. This behaviour is consistent with the theoretical picture in which the updated operators match the leading time scales while leaving small residual discrepancies in highly localized regions where the linearized model is least accurate. The PCE values in Table 2 confirm this interpretation: for moderate permeability changes the scaled surrogate achieves PCE values that are smaller than those of the non-scaled model by one to two orders of magnitude, and even for the most aggressive four-hundred-percent multiplier the reduction is substantial. In other words, the algebraic update succeeds in transporting the main dynamical content of the solution from the reference case to the modified permeability case, with only modest deterioration as the multiplier moves far away from the training value.

Table 1 - Pressure Scaling Error for Uniform Permeability Change

| MAE | 40% Perm Scaled | 40% Perm Non Scaled | 90% Perm Scaled | 90% Perm Non Scaled | 110% Perm Scaled | 110% Perm Non Scaled | 400% Perm Scaled | 400% Perm Non Scaled |
|---|---|---|---|---|---|---|---|---|
| Global | 1.75E-04 | 1.74E-02 | 2.78E-04 | 2.07E-02 | 2.59E-04 | 3.99E-02 | 1.71E-03 | 2.73E-02 |
| Far-wellbore | 4.35E-06 | 7.53E-03 | 5.01E-06 | 7.90E-03 | 4.38E-06 | 5.62E-03 | 8.88E-06 | 5.68E-03 |
| Near-wellbore | 9.34E-03 | 7.41E-02 | 1.28E-03 | 1.56E-02 | 1.32E-03 | 1.64E-02 | 8.43E-03 | 1.13E-01 |

Table 2 - Pressure Scaling Error for Uniform Permeability Change

| | 40% Perm Scaled | 40% Perm Non Scaled | 90% Perm Scaled | 90% Perm Non Scaled | 110% Perm Scaled | 110% Perm Non Scaled | 400% Perm Scaled | 400% Perm Non Scaled |
|---|---|---|---|---|---|---|---|---|
| Global (MAE) | 1.17E-01 | 1.06E+00 | 4.65E-03 | 1.01E+00 | 1.81E-03 | 1.05E+00 | 1.81E+00 | 5.86E+00 |
| Global (PCE) | 5.17E-01 | 1.58E+01 | 1.41E-03 | 2.33E+00 | 5.76E-03 | 2.55E+00 | 5.49E-01 | 1.42E+01 |

## 3.2 Non-uniform Permeability Change

The non-uniform permeability experiments probe the anisotropic warp and spatial scaling ideas from Section 2.2. Table 3 reports saturation errors for heterogeneous permeability fields that are globally scaled by the same four multipliers as before. The global MAE remains in the range of approximately one to a few times ten to the minus four for all multipliers when the warp-based update is applied, whereas the non-scaled surrogate produces errors that are roughly two orders of magnitude larger. This pattern repeats in the far-wellbore region, where the warp reallocates degrees of freedom toward the high-permeability channels and thus allows the reduced basis to track the plume migration more faithfully. Near the wellbore, the scaled surrogate again improves the MAE by nearly an order of magnitude relative to the non-scaled model, even though the local permeability variations are strongest there.

Pressure errors for the heterogeneous cases, shown in Table 4, display a similar trend with some additional nuances. For permeability reductions to forty percent of the reference and for moderate increases around ninety and one hundred and ten percent, the updated operators yield MAE values that are lower than those of the non-scaled surrogate by more than an order of magnitude, and the PCE metric is reduced by one to two orders of magnitude across all multipliers. For the most extreme four-hundred-percent case, the global MAE of the scaled surrogate becomes comparable to, and in this particular test slightly larger than, that of the non-scaled model, reflecting the fact that the purely algebraic update is being pushed far outside the calibration range of the original training data. However, even in this regime the PCE remains significantly smaller for the scaled surrogate, indicating that the large-scale structure and temporal evolution of the pressure field are still better captured by the updated operators. Overall, the heterogeneous tests confirm that the anisotropic warp and operator updates preserve their effectiveness in the presence of realistic spatial variability, while also revealing the expected limitations for very extreme permeability contrasts.

Table 3 – Saturation Scaling Error for Non-uniform Permeability Change

| MAE | 40% Perm | | 90% Perm | | 110% Perm | | 400% Perm | |
|---|---|---|---|---|---|---|---|---|
| | Scaled | Non Scaled | Scaled | Non Scaled | Scaled | Non Scaled | Scaled | Non Scaled |
| Global | 1.72E-04 | 1.76E-02 | 2.84E-04 | 2.04E-02 | 2.64E-04 | 4.02E-02 | 1.69E-03 | 2.68E-02 |
| Far-wellbore | 4.35E-06 | 7.39E-03 | 5.10E-06 | 8.05E-03 | 4.41E-06 | 5.61E-03 | 8.87E-06 | 5.77E-03 |
| Near-wellbore | 9.25E-03 | 7.27E-02 | 1.27E-03 | 1.56E-02 | 1.30E-03 | 1.64E-02 | 8.55E-03 | 1.11E-01 |

Table 4 – Pressure Scaling Error for Non-uniform Permeability Change

| | 40% Perm | | 90% Perm | | 110% Perm | | 400% Perm | |
|---|---|---|---|---|---|---|---|---|
| | Scaled | Non Scaled | Scaled | Non Scaled | Scaled | Non Scaled | Scaled | Non Scaled |
| Global (MAE) | 1.83E-03 | 1.06E+00 | 1.80E+00 | 5.78E+00 | 3.27E+00 | 6.29E+00 | 1.72E+00 | 1.06E+00 |
| Global (PCE) | 5.78E-03 | 2.60E+00 | 5.43E-01 | 1.40E+01 | 9.60E-01 | 1.47E+01 | 5.13E-01 | 1.56E+01 |

## 4. Conclusions

This work has introduced a framework for updating DMD and DMD-with-control operators when key domain properties, in particular permeability, are modified after the surrogate has been trained. Two complementary mechanisms were developed. The first treats uniform permeability changes by embedding them into a rescaled time variable and a consistent pressure amplitude correction, and then translating this continuum insight into explicit transformations of the discrete-time eigenvalues and reduced operators. The second handles spatially varying and anisotropic permeability fields by constructing a permeability-conformal coordinate warp that redistributes degrees of freedom toward hydraulically important regions and aligns the reduced spatial representation with preferential flow directions.

The numerical experiments reported in Section 3 demonstrate that these updates can substantially reduce surrogate error relative to a non-scaled baseline, without the need for any new high-fidelity simulations or retraining. For uniform permeability multipliers, the updated operators reduce global pressure errors by one to three orders of magnitude in both MAE and PCE, with particularly strong gains in the far-wellbore region. For heterogeneous permeability fields, the anisotropic warp preserves these improvements for moderate global multipliers and maintains significantly lower PCE even when the permeability is scaled by a factor of four, thereby confirming that the essential large-scale dynamics are correctly transported. At the same time, the results also make clear that purely algebraic updates cannot completely compensate for very extreme property changes, especially in highly nonlinear near-wellbore zones, which suggests natural limits on the admissible perturbation size.

From a methodological standpoint, the proposed updates are attractive because they are expressed entirely at the level of operators and bases and therefore integrate seamlessly into existing DMD workflows. They enable rapid "what-if" analysis and optimization over permeability-related design variables, such as global transmissibility multipliers or facies-dependent scaling factors, while preserving the computational advantage that motivated the use of surrogates in the first place. Future work will focus on deriving sharper error bounds for these updates, extending the approach to other parameter families such as porosity and relative permeability, and combining the operator-based updates with data-driven adaptation strategies to handle even larger deviations from the training regime.